\documentclass[aps,prd,nofootinbib]{revtex4}
\usepackage[colorlinks=true, pdfstartview=FitV, linkcolor=blue, citecolor=red, urlcolor=magenta]{hyperref}
\usepackage{graphicx}
\usepackage{latexsym}
\usepackage{amsmath}
\usepackage{amsfonts}
\usepackage{amssymb}
\usepackage{verbatim}
\usepackage{dsfont}

\def\t{\tau}

\newcommand{\be}{\begin{equation}}
\newcommand{\ee}{\end{equation}}
\newcommand{\bea}{\begin{eqnarray}}
\newcommand{\eea}{\end{eqnarray}}

\newcommand{\ben}{\begin{eqnarray}}
\newcommand{\een}{\end{eqnarray}}

\begin{document}

\title{Thermal Casimir effect in a classical liquid in a quasi-periodically identified conical spacetime}


\author{$^{1}$K. E. L. de Farias}
\email{klecio.lima@academico.ufpb.br}

\author{ $^{2}$ Azadeh Mohammadi}
\email{azadeh.mohammadi@ufpe.br}

\author{$^{1}$ Herondy F. Santana Mota}
\email{hmota@fisica.ufpb.br}

\affiliation{$^{1}$Departamento de F\' isica, Universidade Federal da Para\' iba,\\  Caixa Postal 5008, Jo\~ ao Pessoa, Para\' iba, Brazil.}
\affiliation{$^{2}$ Departamento de F\'isica, Universidade Federal de Pernambuco,\\
 Av. Prof. Moraes Rego, 1235, Recife - PE - 50670-901, Brazil.}



\begin{abstract}
In this paper, we study the finite-temperature quantum fluctuation of a classical liquid induced by the topology of an effective conical spacetime, as well as by a quasi-periodic boundary condition. The conical spacetime could be either a disclination or a cosmic string. In this context, we consider a phonon field representing quantum excitations of the liquid density, which obeys an effective Klein-Gordon equation with the sound velocity replaced by the light velocity. We obtain closed analytic expressions for the thermal Hadamard function, and consequently, the renormalized mean square density fluctuation of the liquid along with thermodynamics quantities such as internal energy, free energy, total energy, and entropy densities. We also discuss the limiting cases, including low and high-temperature regimes, and the situations in which there are only either the conical spacetime or quasi-periodicity.

\end{abstract}
 \maketitle

\section{Introduction}


\label{sec1}

The quantum field theory at zero temperature is well established and constitutes the basis of the Standard Model of Particle Physics, providing several predictions that have been verified with experiments. However, zero temperature does not generally occur in the universe, and in order to have more realistic models, one needs to include temperature corrections. The presence of temperature modifies the behavior and the properties of the system, as seen for example, in the study of superconductors carried out by the Dutch physicist Heike Kamerlingh Onnes who discovered in 1911 that at low temperatures, the electrical resistance of the material goes to zero \cite{kamerlingh1911resistance}. Nevertheless, a finite-temperature model is usually more complicated than the zero-temperature one, demanding more sophisticated mathematical or even numerical techniques.

On the other hand, the lattice vibration model proposed by Einstein in 1907 and Debye in 1912 \cite{einstein1907plancksche,debye1912theorie} made it possible to connect the elementary vibrations in a solid with the specific heat. Quantically, as the lattice vibrations can be understood as phonon's excitations, heat transfer in this scenario has been studied recently in \cite{Fong:2019mta}, reinforcing that the phonon theory is related in a direct way with temperature. Hence, it becomes natural to explore quantum thermal effects when phonons are included in a system since they represent quantum vibrations in solids and fluids. Although the solid and fluid differentiate themselves by the organization of the atomic lattice, the thermodynamics quantities of the fluid can be interpreted in the same way as those of the solid with the difference of the possible direction of polarization of the sound waves, which in the fluid is only longitudinal and in the solid longitudinal and transversal \cite{lifshitz2013statistical}. Hence, the temperature shows to be of high significance since the phonon's excitations are increased with temperature.

Another component that affects the quantum fluctuation is the topology of the spacetime, where the phonon's modes can propagate. The nontrivial topology of the spacetime induces a modification in the quantum fluctuations of the field just like a boundary condition does \cite{Mostepanenko:1990ceg,Bordag:2009zz}. The cosmic string spacetime, for instance, has a conical topology that is codified by the cosmic string parameter $q>1$. The latter is associated with the linear mass density of the cosmic string, $\mu_0$, by means of the relation $q^{-1}=1-G\mu_0$, where $G$ is Newton's gravitational constant \cite{vilenkin2000cosmic,hindmarsh1995cosmic,allen1990evolution}. In contrast, in the case of a disclination, the cosmic string counterpart in condensate matter, the disclination parameter takes values $q>0$ \cite{Katanaev:1992kh}.

Additionally to the nontrivial topology of the spacetime, the quantum fluctuation of a field can also be modified by an imposed boundary condition \cite{Mostepanenko:1990ceg}. Of particular interest is the quasi-periodic condition in a conical spacetime given, in cylindrical coordinates, by $\Phi(t,r,\varphi,z)=e^{-2\pi ib}\Phi(t,r,\varphi+2\pi/q,z)$, where $0\le b<1$. Note that we recover the known periodic condition for $b=0$, as well as the anti-periodic condition for $b=1/2$. Thereby, the solution of the equation of motion will present the explicit dependence on the parameter $q$, as well as on the parameter $b$. Physically, as argued in recent works, the latter can be interpreted as a control parameter for properties occurring in nanotubes \cite{Saharian:2009ed,klecio2020quantum}.

In the context of finite-temperature effects considering the conical topology of the cosmic string spacetime, Davies and Sahni \cite{Davies:1987th} analyzed the cosmic string immersed in a bath of primordial heat radiation. Moreover, thermal effects have been explored in several other scenarios by using the euclidian action \cite{Martinez:1990sd,Linet:1992qb,Linet:1995ws} and the heat-kernel method \cite{Fursaev:1993qk,Frolov:1995hu}. Recently, the study of finite-temperature effects in the induced current density by considering fermionic and bosonic systems in the cosmic string spacetime has also been studied \cite{Mohammadi:2014lwa,Mohammadi:2015mha}. In the present work, we wish to investigate thermal effects in a scenario where a phonon quantum field is subject to a quasi-periodic condition and whose modes propagate in a spacetime with conical topology, which can be either a cosmic string or a disclination. The phonon modes represent vibrations in a liquid medium in the same way as considered in Ref. \cite{deFarias:2021qdg} at zero temperature, providing a good opportunity to study an analog system in condensed matter, to the Casimir effect in quantum field theory. Here, the thermal Hadamard function is obtained in this configuration and makes it possible to calculate thermalized physical observables such as the liquid density, internal energy, free-energy, and entropy densities. This method introduces temperature corrections in the observables by using the density matrix and the partition function associated with an anti-commutation relation between the field solution and its conjugated counterpart. Hence, the introduction of temperature corrections aims to generalize the results obtained in \cite{deFarias:2021qdg} and complement the ones obtained in Refs. \cite{ford2009fluid,ford2009phononic}.

The paper is organized as follows. In Sec.\ref{sec2} we present the thermalization process by calculating closed and exact analytical expressions for the Hadamard function, as well as for the mean square density fluctuation of the liquid, as a consequence of the imposition of a quasi-periodic condition on the phonon field in the conical structure of a cosmic string (or disclination) spacetime. In Sec.\ref{sec3} we calculate and analyze the influence of the temperature in the energy densities of the system, i.e., internal energy density, free energy density, and the total energy density. In Sec.\ref{sec4} we also calculate the entropy density and discuss its behavior in terms of the temperature. Finally, in Sec.\ref{sec5} we present our conclusions.


\section{Thermal mean square density }
\label{sec2}

The quantum fluctuation study in a fluid is realized by perturbing the fluid mass density $\rho$, with the subsequent quantization giving rise to massless excitations described by phonons. If the perturbation is linear (small), it can be written in the form $\rho'=\rho-\rho_0$, with $\rho_0$ being a constant mean mass density, and $\rho^{\prime}$ the perturbation about $\rho_0$. The small perturbations lead to a linear dispersion relation $\omega=u|k|$, where $u$ is the sound velocity in the liquid. Thereby, the mass density perturbation $\rho'$ is related to the velocity of the fluid $\vec{v}$ by the continuity equation in the following form, \cite{lifshitz2013statistical}
\begin{equation}
\frac{\partial\rho^{\prime}}{\partial t}=\nabla\cdot(\rho\vec{v})\approx-\rho_0\nabla\cdot\vec{v}=-\rho_0\nabla^2\phi,
\label{0.1}
\end{equation}
where second-order terms are neglected. If the fluid is irrotational, one can write the fluid's velocity in terms of a gradient of a massless real scalar field, that is, $\vec{v} = \nabla\phi$. The scalar field, $\phi$, thus, represents the quantum excitations (phonons) of the fluid, which in our case is a liquid. Note that we used this assumption in Eq. \eqref{0.1}.

Following canonical quantization rules, the quantum description of the liquid is reached once we replace the classical hydrodynamics quantities with operators expressed in terms of phonon annihilation and creation operators $\hat{c_k}$ and $\hat{c}^{\dagger}_k$, respectively. They satisfy the following commutation relation $\left[\hat{c}_k,\hat{c}^{\dagger}_{k^\prime}\right]=\delta_{kk^\prime}$, with $\delta_{kk^\prime}$ being either a Kronecker or a Dirac delta depending on whether the set of field modes $k$ is discrete or continuous, respectively.
Hence, by construction, the density perturbation and velocity potential operators should obey the following commutation rule
\begin{equation}
\hat{\phi}(\vec{r})\hat{\rho}^{\prime}(\vec{r}\;^{\prime})-\hat{\rho}^{\prime}(\vec{r}\;^{\prime})\hat{\phi}(\vec{r})=-i\hbar\delta^3(\vec{r}-\vec{r}\;^{\prime}),
\label{1.1}
\end{equation}
where $\delta^3(\vec{r}-\vec{r}\;^{\prime})$ is the Dirac delta function. Note that the relation between the operators is analogous to the field and its momentum conjugate in quantum field theory.
Finally, the density perturbation operator can be expressed in terms of the time derivative of the velocity potential operator as \cite{lifshitz2013statistical, deFarias:2021qdg}
\begin{equation}
\hat{\rho}^{\prime}(t,\vec{r})=-\frac{\rho_0}{u^2}\dot{\hat{\phi}}(t,\vec{r}).
\label{1.4}
\end{equation}
%

In \cite{deFarias:2021qdg}, we have studied modifications on the quantum vacuum fluctuations by considering that the phonon modes propagate in the effective (3+1)-dimensional cosmic string spacetime under a quasi-periodic boundary condition at zero temperature. However, the effect of temperature can be significant in physical quantities. It may modify or generate interesting phenomena, especially in the context of fluids. Here, we will address some of the physical aspects of the finite temperature considering one of the physical systems studied in \cite{deFarias:2021qdg} at zero temperature, namely, the quasi-periodically identified cosmic string spacetime. The investigation here is also valid for disclination.


Let us start with the line element of an effective cosmic string or disclination spacetime in a liquid
\begin{equation}
ds^2=g_{\mu\nu}dx^{\mu}dx^{\nu}=c^2dt^2-dr^2-r^2d\varphi^2-dz^2,
\label{2.0}
\end{equation}
with $r\geq0$, $\varphi\in[0,2\pi/q]$ and $t,z\in(-\infty,+\infty)$ where $q$ encodes the conicity of the background spacetime. When $q=1$, the conical structure disappears, and one recovers the Minkowski spacetime.

The solution of the massless Klein-Gordon equation in a quasi-periodically identified conical spacetime is written as \cite{deFarias:2021qdg}
\begin{equation}
\phi(t,r,\varphi,z)=Ae^{-i\omega_k t}e^{i\nu z}e^{iq(n+b)\varphi}J_{q|n+b|}(\eta r),
\label{2.2}
\end{equation}
which satisfies the condition
\begin{equation}
\Phi(t,r,\varphi,z)=e^{-2\pi ib}\Phi(t,r,\varphi+2\pi/q,z),
\label{2.1}
\end{equation}
with $0\leq b < 1$. Note that for $b=0$ and $b=1/2$ the condition \eqref{2.1} recovers the widely known periodic and anti-periodic conditions, respectively.

The parameter $A$ in the above solution is a normalization constant, $\omega^2_k=u^2(\nu^2+\eta^2)$ is the dispersion relation, $k=(n,\eta,\nu)$ is the set of quantum numbers and $J_{\mu}(x)$ is the Bessel function of first kind.
In the second quantized form, the field operator is written as
\begin{equation}
\hat{\phi}(t,r,\varphi,z)=\sum_{\{k\}}\left[A_k\hat{c}_ke^{-i(\omega_k t-\nu z-q(n+b)\varphi)}+A_k^{*}\hat{c}^{\dagger}_ke^{i(\omega_k t-\nu z-q(n+b)\varphi)}\right]J_{q|n+b|}(\eta r),
\label{2.5}
\end{equation}
with the normalization constant
\begin{equation}
|A_k|=\sqrt{\frac{qu^2\hbar\eta}{2(2\pi)^2\rho_0\omega_k}}.
\label{2.6}
\end{equation}
The symbol
\begin{equation}
\sum_{\{k\}}=\int_{-\infty}^{\infty}\!\!\!\!d\nu\int^{\infty}_{0}\!\!\!\!d\eta\sum_{n=-\infty}^{\infty},
\label{symbol}
\end{equation}
denotes the sum over all quantum numbers (see \cite{deFarias:2021qdg} for a more detailed analysis).
In order to study the behavior of the density fluctuation in a finite temperature scenario, we calculate the thermal Hadamard function, which is fundamental to obtain the physical observables associated with the system, i.e., the renormalized mean square density fluctuation, the free energy density, and the entropy density.
The thermal Hadamard function can be calculated employing \cite{birrell1984quantum}
\begin{equation}
G^{(1)}(x, x^{\prime})=\text{Tr}[\hat{\varrho}(\phi^*(x^{\prime})\phi(x)+\phi(x)\phi^*(x^{\prime}))],
\label{2.7}
\end{equation}
where $x\equiv(\t,r,\varphi,z)$ and $\hat{\varrho}$ is the density matrix,
\begin{equation}
\hat{\varrho}=Z^{-1}e^{-\beta \hat{H}},
\label{2.7.1}
\end{equation}
with $\beta=\frac{1}{k_BT}$ and $\hat{H}$ is the Hamiltonian operator. The partition function $Z$ is described by
\begin{equation}
Z=\text{Tr}[e^{-\beta \hat{H}}].
\label{2.7.2}
\end{equation}
As we need to deal with a scalar field, we take
\begin{equation}
\text{Tr}[\hat{\varrho}\hat{a}^+_{\sigma}\hat{a}_{\sigma^{\prime}}]=\frac{\delta_{\sigma\sigma^{\prime}}}{e^{\beta\hslash\omega_k}-1},
\label{2.7.3}
\end{equation}
compatible with the Bose-Einstein statistics.
Substituting the field operator \eqref{2.5} in Eq. \eqref{2.7} and considering the relation in \eqref{2.7.3}, we obtain
\begin{eqnarray}
G^{(1)}(x, x^{\prime})&=&\sum_{\{k\}}\frac{qu^2\hbar\eta}{8\pi^2\rho_0\omega_k}e^{i[\nu\Delta z+q(n+b)\Delta\varphi]}J_{q|n+b|}(r\eta)J_{q|n+b|}(r^{\prime}\eta)\nonumber\\
\ &\ &\times\left[\left(e^{-i\omega_k\Delta t}+e^{i\omega_k\Delta t}\right)+2\left(\frac{e^{-i\omega_k\Delta t}+e^{i\omega_k\Delta t}}{e^{\beta\hslash\omega_k}-1}\right)\right].
\label{2.8}
\end{eqnarray}
One should note that the thermal Hadamard function above can be divided into two parts considering the terms in brackets on the r.h.s. of \eqref{2.8}, that is,
\begin{eqnarray}
G^{(1)}(x, x^{\prime})=G^{(1)}_0(x, x^{\prime})+G^{(1)}_T(x, x^{\prime}).
\label{2.9}
\end{eqnarray}
The first term on the r.h.s. refers to the zero temperature two-point Hadamard function. The second term is the two-point thermal Hadamard function. The whole temperature effect is encoded in the latter which is the part we are interested in here since the non-thermal part has been already investigated in \cite{deFarias:2021qdg}. Therefore, we can focus on the thermal part of the two-point Hadamard function. From Eq. \eqref{2.8}, it is given by
\begin{eqnarray}
G^{(1)}_T(x, x^{\prime})=\sum_{\{k\}}\frac{qu^2\hbar\eta}{4\pi^2\rho_0\omega_k}\frac{(e^{-i\omega_k\Delta t}+e^{i\omega_k\Delta t})}{e^{\beta\hslash\omega_k}-1}e^{i[\nu\Delta z+q(n+b)\Delta\varphi]}J_{q|n+b|}(r\eta)J_{q|n+b|}(r^{\prime}\eta).
\label{2.9.1}
\end{eqnarray}
Furthermore, Eq. \eqref{2.9.1} can be worked out by using the identity
\begin{equation}
(e^{y}-1)^{-1}=\sum_{j=1}^{\infty}e^{-jy}.
\end{equation}
Then, by making the wick rotation $i\Delta t=\Delta\tau$, taking the following recurrence relation
\begin{equation}
\frac{e^{-\omega_k(\delta\Delta\tau+\beta\hslash j)}}{\omega_k}=\frac{2}{\sqrt{\pi}}\int_{0}^{\infty}ds e^{-s^2\omega_k^2-(\delta\Delta\tau+\hslash\beta j)^2/4s^2},
\label{2.10}
\end{equation}
where $\delta=\pm1$ refers to positive and negative frequencies, and considering Eq. (21) in Ref. \cite{bragancca2019vacuum}, it is possible to put the thermal Hadamard function \eqref{2.9.1} in the form
\begin{eqnarray}
G^{(1)}_T(x,x^{\prime})&=&\frac{qu^2\hbar}{4\pi^2\rho_0}\frac{e^{iqb\Delta\varphi}}{\sqrt{\pi}}\sum_{\delta=+,-}\sum_{j=1}^{\infty}\sum_{n=-\infty}^{\infty}e^{inq\Delta\varphi}\int_{-\infty}^{\infty}d\nu e^{i\nu\Delta z}\int_{0}^{\infty}\frac{ds}{s^2}e^{-(su)^2\nu^2-\frac{\Delta\zeta^2}{4(su)^2}}I_{q|n+b|}(rr^{\prime}/2(su)^2),\nonumber\\
\label{2.10}
\end{eqnarray}
where $\Delta\zeta^2=(\delta\Delta\tau+\beta\hslash j)^2u^2+r^2+r^{\prime2}$. Finally, thanks to the summation formula (20) in Ref. \cite{deFarias:2021qdg} (see also (A.10) in Ref. \cite{de2015vacuum}) the sum in the parameter $n$ can be performed and, consequently, Eq. \eqref{2.10} becomes
\begin{eqnarray}
G^{(1)}_T(x,x^{\prime})&=&\frac{u\hbar e^{iqb\Delta\varphi}}{4\pi^2\rho_0rr^{\prime}}\sum_{\delta=+,-}\sum_{j=1}^{\infty}\left\{\sum_n e^{ib(2\pi n-q\Delta\varphi)}\frac{1}{\sigma_n}\right.\nonumber\\
\ &\ &\left.-\frac{q}{2\pi i}\sum_{\ell=+,-}\ell e^{i\ell qb\pi}\int_{0}^{\infty}dy\frac{1}{\sigma_y}\frac{\cosh[qy(1-b)]-\cosh(qb y) e^{-iq(\ell\pi+\Delta\varphi)}}{\cosh(qy)-\cos[q(\Delta\varphi+\ell\pi)]}\right\},
\label{2.11}
\end{eqnarray}
with
\begin{eqnarray}
\sigma_n&=&\frac{\Delta\zeta^2}{rr^{\prime}}+\frac{\Delta z^2}{2rr^{\prime}}-\cos(2\pi n/q-\Delta\varphi),\nonumber\\
\sigma_y&=&\frac{\Delta\zeta^2}{rr^{\prime}}+\frac{\Delta z^2}{2rr^{\prime}}+\cosh y.
\end{eqnarray}
Note that the sum in $n$ is restricted to the interval \cite{de2015vacuum, deFarias:2021qdg}
\begin{equation}
-\frac{q}{2}+\frac{\Delta\varphi}{\varphi_0}\leq n\leq\frac{q}{2}+\frac{\Delta\varphi}{\varphi_0}.
\end{equation}
Therefore, the temperature effect is encoded in $\sigma_n$  and $\sigma_y$ through $\Delta\zeta$ defined below Eq. \eqref{2.10}.

For completeness, we can follow a similar procedure and find the zero-temperature part of \eqref{2.9}, which is obtained as
\begin{eqnarray}
G_{0}^{(1)}(x,x^{\prime})&=&\frac{qu\hbar e^{iqb\Delta\varphi}}{8\pi^2\rho_0rr^{\prime}}\sum_{\delta=+,-}\left\{\frac{1}{q}\sum_n e^{ib(2\pi n-q\Delta\varphi)}\frac{1}{\sigma_n}\right.\nonumber\\
\ &\ &\left.-\frac{1}{2\pi i}\sum_{\ell=+,-}\ell e^{i\ell qb\pi}\int_{0}^{\infty}dy\frac{1}{\sigma_y}\frac{\cosh[qy(1-b)]-\cosh(qb y) e^{-iq(\ell\pi+\Delta\varphi)}}{\cosh(qy)-\cos[q(\Delta\varphi+\ell\pi)]}\right\}.
\label{2.11.1}
\end{eqnarray}
The non-thermal two-point function in Eq. \eqref{2.11.1} agrees with the one found in the Ref. \cite{deFarias:2021qdg}, as expected. Note that by considering $n=0$ in Eq. \eqref{2.11.1}, we obtain the Minkowski contribution, which should be subtracted in the renormalization process since it is divergent.
 It is worth mentioning that the renormalization process is responsible for removing the divergent term of the expression, which arises by taking the coincidence limit $x^{\prime}\to x$ (for the detailed process, see \cite{deFarias:2021qdg}). However, in contrast with the zero-temperature part, the $n=0$ term in Eq. \eqref{2.11} is not divergent and may give a relevant contribution to the physical observables, as we shall see.

Since we have found the thermal Hadamard function \eqref{2.11} in closed-form, we are now able to calculate the relevant physical observables. This is done from now on.

\subsection{Mean Square Density}
\label{subsec}

Having the finite temperature two-point Hadamard function at hand, we can calculate one of the most important physical quantities in the context of the quantum excitations of a classical fluid, the mean square density fluctuation, in the effective conical spacetime \eqref{2.0} imposing the quasi-periodic condition \eqref{2.1}. It is straightforward to realize that because of the linear nature of the density operator $\hat{\rho}$ with the annihilation $\hat{c}$ and creation $\hat{c}^{\dagger}$ operators, the vacuum expectation value $\langle\hat{\rho}\rangle$ vanishes \cite{deFarias:2021qdg}. Consequently, the mean square density $\langle\hat{\rho}^2\rangle$ should be analyzed, but first, it is convenient to rewrite it in terms of the thermal Hadamard function. To compute the mean square density fluctuation, we start by considering the product
\begin{eqnarray}
\hat{\rho}(x)\hat{\rho}(x')&=&\frac{\rho_0^2}{u^4}\frac{\partial^2}{\partial t\partial t^{\prime}}[\hat{\phi}(x)\hat{\phi}(x')].
\label{2.14}
\end{eqnarray}
The ensemble average of an arbitrary operator $\hat A$  submitted to temperature $T=(k_B\beta)^{-1}$ is given by \cite{birrell1984quantum}
\begin{equation}
\langle \hat A\rangle_{\beta}=\sum_i\varrho_i\langle\psi_i|\hat A|\psi_i\rangle,
\label{2.14.0}
\end{equation}
where $\varrho_i=\langle\psi_i|\hat{\varrho}|\psi_i\rangle$, with $\hat{\varrho}$ being the density matrix described by  Eq. \eqref{2.7.1}. One can express \eqref{2.14.0} in the following form:
\begin{eqnarray}
\langle \hat A\rangle_{\beta}&=&\sum_{i,j}\langle\psi_i|\hat \rho|\psi_j\rangle\langle\psi_j|\hat A|\psi_i\rangle\nonumber\\
\ &=&\sum_i\langle\psi_i|\rho \hat A|\psi_i\rangle={\rm Tr}(\hat \rho \hat A).
\label{2.14.1}
\end{eqnarray}
Therefore, considering the thermal average of \eqref{2.14}, the mean square density at finite temperature can be written as follows
\begin{eqnarray}
\langle\hat{\rho}(x)\hat{\rho}(x')\rangle&=&\frac{\rho_0^2}{u^4}\frac{\partial^2}{\partial t\partial t^{\prime}}\frac{1}{2}G^{(1)}(x,x^{\prime})\nonumber\\
\ &=&\frac{\rho_0^2}{u^4}\frac{\partial^2}{\partial t\partial t^{\prime}}\frac{1}{2}\left(G^{(1)}_{\rm T}(x,x^{\prime})+G^{(1)}_0(x,x^{\prime})\right)\nonumber\\
\ &=&\langle\hat{\rho}(x)\hat{\rho}(x')\rangle_{\rm T} + \langle\hat{\rho}(x)\hat{\rho}(x')\rangle_0.
\label{2.15}
\end{eqnarray}
As we did before for the two-point Hadamard function, the mean square density can be divided into two parts, the zero and non-zero temperature contributions. The zero-temperature part has already been calculated in \cite{deFarias:2021qdg}. Therefore, we will focus on the non-zero temperature contribution. Taking the coincidence limit $x^{\prime}\rightarrow x$, we can find a closed-form for the thermal part of the mean square density, i.e.,
\begin{eqnarray}
\langle\rho^2\rangle_{\rm T}&=&\frac{\rho_0^2}{u^4}\lim_{x^{\prime}\rightarrow x}\frac{\partial^2}{\partial t\partial t^{\prime}}\frac{1}{2}G^{(1)}_{\rm T}(x,x^{\prime})\nonumber\\
\ &=&\frac{\hbar\rho_0}{\pi^2u}\sum_{j=1}^{\infty}\left\{ \frac{3}{(u\beta\hslash j)^4} + \sum_{n=1}^{[q/2]}\!^* \, 2\cos(2b\pi n)\right.
F_j(s_n)\left.-\frac{q}{\pi}\int_0^{\infty}dy{M(y,b,q)}F_j(s_y)\right\},
\label{2.16}
\end{eqnarray}
wherein we have defined $s_n=\sin(\pi n/q)$, $s_y=\cosh(y/2)$,
\begin{equation}
M(y,b,q)=\frac{\cosh(qb y) \sin[q\pi(1-b)]+\cosh[qy(1-b)]\sin(qb\pi)}{\cosh(qy)-\cos(q\pi)},
\end{equation}
and
\begin{equation}
F_j(s)=\frac{3 (u\beta\hslash j)^2 - (2rs)^2}{\left[(u\beta\hslash j)^2+(2rs)^2\right]^3}.
\end{equation}
The parameter $s$ should be replaced by $s_n$ for the second term on the r.h.s of \eqref{2.16} and $s_y$ for the third term. It is worth highlighting that, the square bracket in $[q/2]$ denotes the floor function and the sign $(*)$ means that for an integer $q$, the sum in $n$ must be replaced by
\begin{equation}
\sum_{n=1}^{[q/2]}\rightarrow\frac{1}{2}\sum_{n=1}^{q-1}.
\end{equation}
The first term on the right-hand side of \eqref{2.16} arises from $n=0$ which corresponds to the scalar black-body radiation originating from the Minkowski contribution at finite temperature. By performing the sum in $j
$ in this term, it may be expressed as
\begin{equation}
\langle\rho^2\rangle_{\rm T}^{\rm rad}=\frac{\pi^2\rho_0}{30\hslash^3\beta^4u^5}.
\label{2.16.0}
\end{equation}
We can now define the renormalized thermal contribution to the mean square density fluctuation as
\begin{equation}
\langle\rho^2\rangle_{\rm T}^{\rm ren}=\langle\rho^2\rangle_{\rm T}-\langle\rho^2\rangle_{\rm T}^{\rm rad},
\label{dif}
\end{equation}
where the Minkowski contribution \eqref{2.16.0} is removed. From now on, every renormalized thermal component means that the radiation term has been removed. Besides the fact that it is a general procedure to remove the Minkowski contribution to obtain renormalized quantities, in the case of thermal expressions, it is also necessary to provide the correct classical behavior at high temperatures for the free energy density and entropy density.

Note that, by taking the limit $\beta\to\infty$ (i.e. $T\to 0$), the thermal part of \eqref{dif} vanishes, which means that the mean square density fluctuation is reduced to the vacuum contribution at zero temperature in Eq. \eqref{2.15}, $\langle\rho^2\rangle\to\langle\rho^2\rangle_0$ as expected. The zero-temperature contribution is given by
\begin{eqnarray}
\langle\rho^2\rangle_{0}^{\rm ren}&=&\frac{\rho_0^2}{u^4}\lim_{x^{\prime}\rightarrow x}\frac{\partial^2}{\partial t\partial t^{\prime}}G^{(1)}_{\rm reg}(x,x^{\prime})\nonumber\\
\ &=&-\frac{\hbar\rho_0}{32\pi^2ur^4}\left\{2\sum_{n=1}^{[q/2]}\!^* \frac{\cos(2b\pi n)}{\sin^4(\pi n/q)}-\frac{q}{\pi}\int_0^{\infty}dy\frac{M(y,b,q)}{\cosh^4(y/2)}\right\},
\label{2.16.1}
\end{eqnarray}
where $G^{(1)}_{\rm reg}$ means $G^{(1)}_0$ without the Minkowski contribution obtained by $n=0$. It is easy to verify that this term matches the one in \cite{deFarias:2021qdg}.

The sum in $j$ present in Eq. \eqref{2.16} can be performed by using the recurrence formula
\begin{equation}
\sum_{m=1}^{\infty}\frac{ \left(3 m^2-w^2\right)}{\left(m^2+w^2\right)^3} = \frac{\left[1-\pi ^3 w^3 \coth (\pi  w) \text{csch}^2(\pi  w)\right]}{2w^4}.
\end{equation}
Hence, Eq. \eqref{dif} leads to the following result
\begin{eqnarray}
\langle\rho^2\rangle_{\rm T}^{\rm ren}&=&\frac{\hbar\rho_0}{32\pi^2r^4u}\left\{\sum_{n=1}^{[q/2]}\!^* \, \right.
2\cos(2b\pi n)F(s_n)\left.-\frac{q}{\pi}\int_0^{\infty}dy
M(y,b,q) F(s_y)\right\},
\label{2.16.2}
\end{eqnarray}
where the dimensionless parameter $\gamma$ is defined as $\gamma=\frac{r}{\hslash u\beta}$ and
\begin{eqnarray}
F(s)=\sum_{j=1}^{\infty} F_j(s)=\frac {1}{s^4}\left[1-(2\pi\gamma s)^3\coth\left(2\pi\gamma s\right)\text{csch}^2\left(2\pi\gamma s\right)\right].
\end{eqnarray}
In Eq. \eqref{2.16.2}, both the finite sum in $n$ and the integral over $y$ can only be calculated adopting numerical analysis. Note that the mean square density fluctuation is proportional to $r^{-4}$, which means it diverges on the string, $r\to 0$, and goes to zero far from it, in the limit  $r\to\infty$.
\begin{figure}[h]
  \includegraphics[scale=0.4]{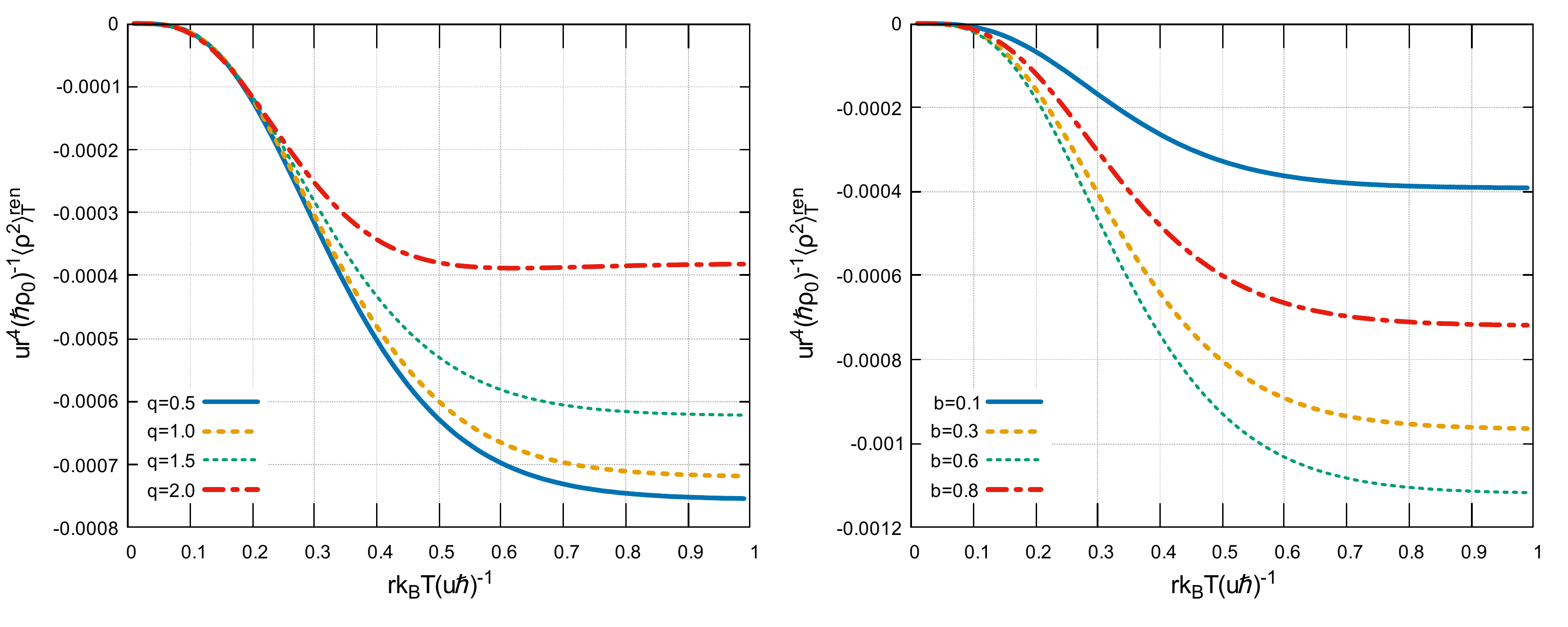}
  \caption{
  Mean square density fluctuation for $b=0.8$ (left panel), and $q=1$ (right panel) as a function of temperature.}
  \label{fig1}
\end{figure}
For large temperatures, $\beta \ll 1$, the function $F(s)$ can be approximated by
\begin{eqnarray}
F(s)\approx \frac {1}{s^4}\left[1-\frac{4(2\pi\gamma s)^3}{e^{4\pi\gamma s}}\right].
\end{eqnarray}
Thus, the mean square density fluctuation goes to
\begin{eqnarray}
\langle\rho^2\rangle_{\rm T\to \infty}^{\rm ren}\approx\frac{\hbar\rho_0}{32\pi^2r^4u}\left\{\sum_{n=1}^{[q/2]}\!^*
\, \frac{ 2\cos(2b\pi n)}{s_n^4}-\frac{q}{\pi}\int_0^{\infty}dy
\frac{M(y,b,q)}{s_y^4}\right\}.
\label{2.16.3}
\end{eqnarray}
One should note that the mean square density above is independent of temperature in the high-temperature limit, $T \to \infty$. This can clearly be observed in the plots of Fig.\ref{fig1}. On the other hand, the plots also show that Eq. \eqref{2.16.2} tends to zero when $T \to 0$, as expected.

Furthermore, we can consider two important particular cases, the quasi-periodically idendified Minkowski spacetime contribution $(q=1)$ and the case of a conical spacetime only $(b=0)$. For $b=0$ the function $M(y,b,q)$ becomes
\begin{eqnarray}
M(y,0,q)=\frac{\sin(\pi q)}{\cosh(q y)-\cos(\pi q)},
\label{2.17}
  \end{eqnarray}
and the expression in \eqref{2.16.2} provides the conical spacetime contribution only. Therefore, in this case, for $q\in\mathds{Z}$, it is straightforward to see that the integral contribution in Eq. \eqref{2.16.2} is null, and only the summation term remains for $q\geq 2$. For $q<2$, on the other hand, there is no summation contribution to the mean square density \eqref{2.16.2}. Of course, in the case of $q=1$ (no conical structure) and $b=0$, the mean square density fluctuation averages to zero. Conversely, for $q=1$, the only contribution comes from the quasi-periodically identified Minkowski spacetime. In this case, we have $M(y,b,1)$, and the only nonzero part of Eq. \eqref{2.16.2} lies in the integral over $y$. This analysis remains the same for all other observables obtained below.


\section{Internal energy, free energy and total energy of the system}
\label{sec3}

Another important physical quantity is the energy of the system. In this sense, the energy of a liquid can be taken as the vacuum expectation value of the Hamiltonian operator written in terms of the massless scalar field operator $\hat{\phi}$ as  \cite{lifshitz2013statistical}
%
\begin{eqnarray}
\hat{H}&=&\int d^3x\left[\frac{1}{2}\hat{v}\cdot\rho\hat{v}+\hat{\rho}\frac{u^2}{2\rho_0}\hat{\rho}\right]\nonumber\\
\ &=&\int d^3x\frac{1}{2}\rho_0\left[\nabla\hat{\phi}\cdot\nabla\hat{\phi}+\frac{1}{u^2}\partial_t\hat{\phi}\partial_t\hat{\phi}\right].
\label{3.1}
\end{eqnarray}
As the system has been considered in an infinite volume, that is, the whole conical spacetime, the integral \eqref{3.1} diverges. However, it is possible to define the Hamiltonian density operator $\hat{\mathcal{H}}$ as being the integrand of Eq. \eqref{3.1}, i.e,
\begin{eqnarray}
\hat{\mathcal{H}}=\frac{\rho_0}{2}\left[\nabla\hat{\phi}\cdot\nabla\hat{\phi} + \frac{1}{u^2}\partial_t\hat{\phi}\partial_t\hat{\phi}\right],
\label{3.2}
\end{eqnarray}
where the first term on the r.h.s is associated with the kinetic energy density operator and the second term with the internal energy density operator \cite{lifshitz2013statistical, ford2009fluid,ford2009phononic}. The mean value of the Hamiltonian density operator can be calculated, using the Hadamard function \eqref{2.8}, in the following form
\begin{eqnarray}
\langle \mathcal{H}\rangle=\lim_{x^{\prime}\to x}\frac{\rho_0}{2}\left[\partial_i\partial_{i^{\prime}}G^{(1)}(x,x^{\prime}) + \frac{1}{u^2}\partial_t\partial_{t^{\prime}}G^{(1)}(x,x^{\prime})\right],
\label{3.3}
\end{eqnarray}
where one needs to take the coincidence limit $x^{\prime}\to x$, after subtracting the Minkowski contributions.
As before, Eq. \eqref{3.3} may be divided into two parts,
\begin{equation}
\langle \mathcal{H}\rangle=\langle \mathcal{H}\rangle_T+\langle \mathcal{H}\rangle_0,
\label{3.4}
\end{equation}
representing the energy density due to the presence of temperature $\langle \mathcal{H}\rangle_T$, and the zero-temperature contribution $\langle\mathcal{H}\rangle_0$. The thermal contribution $\langle \mathcal{H}\rangle_T$
can be written as
\begin{eqnarray}
\langle \mathcal{H}\rangle_{\rm T}=\lim_{x^{\prime}\to x}\frac{\rho_0}{2}\left[\frac{1}{u^2}\partial_t\partial_{t^{\prime}}+\partial_r\partial_{r^{\prime}}+\frac{1}{rr^{\prime}}\partial_{\varphi}\partial_{\varphi^{\prime}}+\partial_z\partial_{z^{\prime}} \right]G^{(1)}_T(x,x^{\prime}).
\label{3.5}
\end{eqnarray}
%
Before proceeding to calculate the system's total energy, it is worth exploring the relevant physical quantities attributed to the internal energy, that is, the free energy density of the fluid and the entropy density of the system at finite temperature. We will also explore the impact of the boundary condition and the nontrivial topology on the quantities mentioned above.

\subsection{Internal and Free energy densities}
The internal energy density is identified as being the second term on the r.h.s of the Hamiltonian operator \eqref{3.1} \cite{lifshitz2013statistical, ford2009fluid,ford2009phononic}, that is,
\begin{eqnarray}
\mathcal{U}&=&\lim_{x^{\prime}\to x}\frac{\rho_0}{u^2}\partial_t\partial_{t^{\prime}}\frac{1}{2}G^{(1)}(x,x^{\prime})=\frac{u^2}{\rho_0}\langle\rho^2\rangle.
\label{3.6}
\end{eqnarray}
Note that in the above expression, there are contributions due to the zero-temperature mean square density \eqref{2.16.1} and its thermal correction in Eq. \eqref{2.16}. It is easy to verify that the black-body radiation contribution in \eqref{2.16} to the internal energy above agrees with the one discussed in \cite{lifshitz2013statistical}{\footnote{Chapter 22}}, as it should.

\begin{figure}[h]
  \includegraphics[scale=0.4]{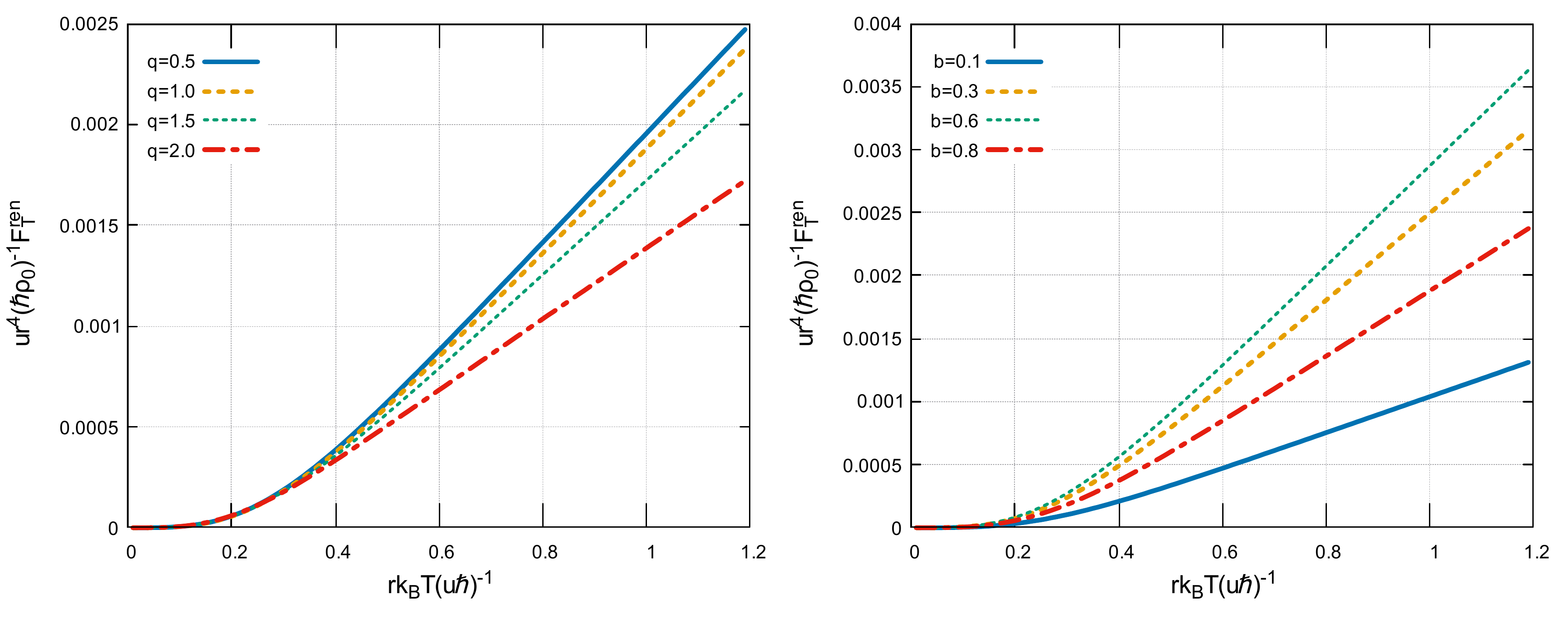}
  \caption{
Free energy density for $b=0.8$ (left panel), and $q=1$ (right panel) as a function of temperature.}
  \label{fig3}
\end{figure}

Once there is an analytical form for the internal energy density, it is possible to calculate the free energy density associated with the system by considering the thermodynamical definition
\begin{equation}
\mathcal{U}=-T^2\frac{\partial}{\partial T}\left(\frac{\mathcal{F}}{T}\right),
\label{3.6.1}
\end{equation}
with $\mathcal{F}$ being the free energy density. Integrating the above equation gives the free energy density up to a temperature-independent term.
In this sense, in terms of the mean square density fluctuation, we have
\begin{eqnarray}
\mathcal{F}=-T\frac{u^2}{\rho_0}\int dT\frac{1}{T^2}\langle\rho^2\rangle=-T\frac{u^2}{\rho_0}\int dT\frac{1}{T^2}\langle\rho^2\rangle_{\rm T} +\frac{u^2}{\rho_0}\langle\rho^2\rangle_0 + C,
\label{3.6.2}
\end{eqnarray}
where $C$ is the temperature-independent integration constant. We will show shortly that the constant should be zero. As our interest lies in the temperature effects (first term on the r.h.s. of \eqref{3.6.2}), the nonthermal contribution will be discarded for the moment. Therefore, the thermal free energy density is given by
\begin{equation}
\mathcal{F}_{\rm T}=-\frac{\hslash u}{\pi^2}\sum_{j=1}^{\infty}\left\{\frac{1}{(u\hslash\beta j)^4}+\frac{\gamma^4}{r^4}\left[\sum^{[q/2]}_{n=1}\!^*2\cos(2b\pi n) h_j(s_n)-\frac{q}{\pi}\int_0^{\infty}dy M(y,b,q)h_j(s_y)\right]\right\},
\label{3.6.3}
\end{equation}
with
\begin{equation}
h_j(s)=\frac{1}{(j^2+4\gamma^2s^2)^2}\, .
\label{3.6.33}
\end{equation}
Performing the sum in $j$, we obtain the following form for free energy density
\begin{eqnarray}
\mathcal{F}_{\rm T}&=&\frac{\hslash u}{\pi^2}\left\{-\frac{\pi^4}{90(u\hslash\beta)^4}-\frac{\gamma^4}{r^4}\left[\sum^{[q/2]}_{n=1}\!^*2\cos(2b\pi n)h(s_n)\right.\right. \left.\left.- \frac{q}{\pi}\int_0^{\infty}dy M(y,b,q)h(s_y)\right]\right\},
\label{3.6.4}
\end{eqnarray}
where
\begin{eqnarray}
h(s)=\sum_{j=1}^{\infty} h_j(s)=\frac{1}{32\gamma^4 s^4}\left[1-\pi\gamma s\left(\coth(2\pi\gamma s)+2\pi s\gamma\text{csch}^2(2\pi\gamma s)\right)\right].
\end{eqnarray}
In the limit, $T\to 0$, the thermal free energy density goes to zero, as expected. The first term on the r.h.s of Eq. \eqref{3.6.4} is the scalar black-body radiation term (thermal Minkowski contribution). For the consistency check, one can verify that this term matches the one in Ref. \cite{lifshitz2013statistical} since it is independent of the boundary and the background imposed on the system. The second and third terms in Eq. \eqref{3.6.4} arise from the quasi-periodic boundary condition and conical spacetime. The renormalized thermal free energy density as a function of temperature for several values of quasi-periodicity parameter $b$ and conicity parameter $q$ is shown in Fig.\ref{fig3}. The plots also show that \eqref{3.6.4} goes to zero as $T\to 0$.

It is worth highlighting that, in contrast with the low-temperature regime, the free energy density shows a linear behavior in the high-temperature limit. To see this better, one can take the asymptotic limit of the hyperbolic functions in Eq. \eqref{3.6.4}, which results in
\begin{eqnarray}
\mathcal{F}_{\rm T}^{\rm ren}&\approx&\frac{\hslash u\gamma}{32\pi r^4}\left\{-\sum^{[q/2]}_{n=1}\!^*2\frac{\cos(2b\pi n)}{s_n^3} + \frac{q}{\pi}\int_0^{\infty}dy \frac{M(y,b,q)}{s_y^3}\right\},
\label{3.6.5}
\end{eqnarray}
where $\gamma\propto k_BT$. The linear behavior proportional to $k_{\rm B}T$, at high $T$ is predicted by the classical theory and it is recovered only if we remove the black-body radiation contribution from \eqref{3.6.4}, in the first term on the r.h.s. This is clearly shown in Fig.\ref{fig3}.


\subsection{Total Energy Density}

Finally, let us study the total energy density of a liquid producing vibrations in the form of quantized sound waves. As already mentioned, the total energy density is composed of the kinetic energy density and internal energy density, the last one given by Eq. \eqref{3.6}. Thus, in order to obtain the total energy density from Eq. \eqref{3.5}, we need to calculate the kinetic energy density given by the spatial derivative terms.
The components $r$ and $z$ are easily obtained by the same process as the one for the temporal component. However, to find the $\varphi$-component, we should use Eq. \eqref{2.10}, and consequently take the derivatives. This leads to
\begin{eqnarray}
\lim_{x^{\prime}\to x}\partial_{\varphi}\partial_{\varphi^{\prime}}G^{(1)}_T(x,x^{\prime})&=& \lim_{\varphi^{\prime}\to \varphi}\frac{qu\hbar}{2\pi^2\rho_0}\frac{e^{iqb\Delta\varphi}}{\sqrt{\pi}}\sum_{\delta=+,-}\sum_{j=1}^{\infty}\int_{-\infty}^{\infty}d\nu e^{i\nu\Delta z}\int_{0}^{\infty}\frac{ds}{s^2}e^{-(su)^2\nu^2-\frac{\Delta\zeta^2}{4(su)^2}}\nonumber\\
\ &\ &\sum_{n=-\infty}^{\infty}e^{inq\Delta\varphi}q^2(n+b)^2I_{q|n+b|}(rr^{\prime}/2(su)^2).
\label{3.7}
\end{eqnarray}
To further proceed, let us make use, in the above result, of the recurrence expression \cite{BezerradeMello:2011sm}
\begin{eqnarray}
q^2(n+b)^2I_{q|n+b|}(w)=\left[w^2\frac{d^2}{dw^2}+w\frac{d}{dw}-w^2\right]I_{q|n+b|}(w).
\label{3.8}
\end{eqnarray}
After employing again the summation formula (20) in Ref. \cite{deFarias:2021qdg} for $\Delta\varphi =0$, we are able to simplify Eq. \eqref{3.7}. Hence, we find that the kinetic part takes the following form
\begin{eqnarray}
\lim_{x^{\prime}\to x}\frac{\rho_0}{2}\nabla\cdot\nabla^{\prime}G^{(1)}(x,x^{\prime}) &=&
\frac{u\hslash}{\pi^2\rho_0}\sum_{j=1}^{\infty}\left\{\frac{3}{u\hslash\beta j}+ \frac{\gamma^4}{r^4} \left[\sum_{n=1}^{[q/2]}\!^*2\cos(2\pi bn)\frac{[j^2(3-4s_n^2)-2\gamma^2(1-6s_n^2)+2\gamma^2\cos(\frac{4\pi n}{q})]}{(j^2+4\gamma^2s_n^2)^3} \right.\right.\nonumber\\
\ &\ &-\frac{q}{\pi}\int_0^{\infty}dy M(y,b,q)\left.\left.\frac{[j^2(3-4s_y^2)-2\gamma^2(1-6s_y^2)+2\gamma^2\cosh(2y)]}{(j^2+4\gamma^2s_y^2)^3}\right]\right\}.
\label{3.9}
\end{eqnarray}
Finally, by summing the internal energy density \eqref{3.6} with the kinetic energy density \eqref{3.9} the total energy density is obtained as
\begin{eqnarray}
\langle\mathcal{H}\rangle_T&=&\frac{2u\hslash}{\pi^2}\sum_{j=1}^{\infty}\left\{\frac{3}{(u\hslash j\beta)^4}+\frac{\gamma^4}{r^4}\left[\sum_{n=1}^{[q/2]}\!^*2\cos(2\pi bn) \chi_j(s_n)-\frac{q}{\pi}\int_0^{\infty}dy M(y,b,q)\chi_j(s_y)\right]\right\},
\label{3.10}
\end{eqnarray}
with
\begin{eqnarray}
\chi_j(s)=\frac{[j^2(3-2s^2)+8\gamma^2s^4-4\gamma^2s^2]}{[j^2+4\gamma^2s^2]^3}.
\end{eqnarray}
Note that we have neglected the zero-temperature contribution in Eq. \eqref{3.10}. In the above expression, performing the summation in $j$, we are led to the following expression
\begin{eqnarray}
\langle\mathcal{H}\rangle_T&=& \frac{\pi^2}{15(u\hslash)^3 \beta^4}\nonumber\\
\ &\ &+\frac{2\gamma^4u\hslash}{\pi^2 r^4}\left\{\sum_{n=1}^{[q/2]}\!^*2\cos(2\pi bn)
\chi(s_n)-\frac{q}{\pi}\int_0^{\infty}dy M(y,b,q) \chi(s_y)\right\},
\label{3.11}
\end{eqnarray}
where
\begin{eqnarray}
\chi(s)&=&\sum_{j=1}^{\infty}\chi_j(s)\nonumber\\
&=&\frac{1}{32\gamma^4}\left[\frac{(1-2s^2)}{s^4} + \frac{\pi\gamma}{s}\coth\left(2\pi\gamma s\right) + 2\pi^2\gamma^2\text{csch}^2\left(2\pi\gamma s\right)\left(1- \frac{4\pi\gamma(1-s^2)}{s}\coth\left(2\pi\gamma s\right)\right)\right].
\end{eqnarray}

For completeness we can also exhibit, by following the same steps as above, the zero-temperature contribution to the energy density. It is given by
\begin{eqnarray}
\langle\mathcal{H}\rangle_0=-\frac{u\hslash}{16\pi^2r^4}\left\{\sum_n^{[q/2]}\!^*\,  2\cos(2\pi bn)\frac{[1-2s_n^2]}{s_n^4}-\frac{q}{\pi}\int_0^{\infty}M(y,b,q)\frac{[1-2s_y^2]}{s_y^4}\right\}.
\label{3.12}
\end{eqnarray}
In the context of liquids, the total energy density at zero temperature has not been calculated previously in the literature. Note that Eq. \eqref{3.12} is compatible with the energy density of a massless quantum scalar field studied in the Ref. \cite{klecio2020quantum}.

\begin{figure}[h]
  \includegraphics[scale=0.4]{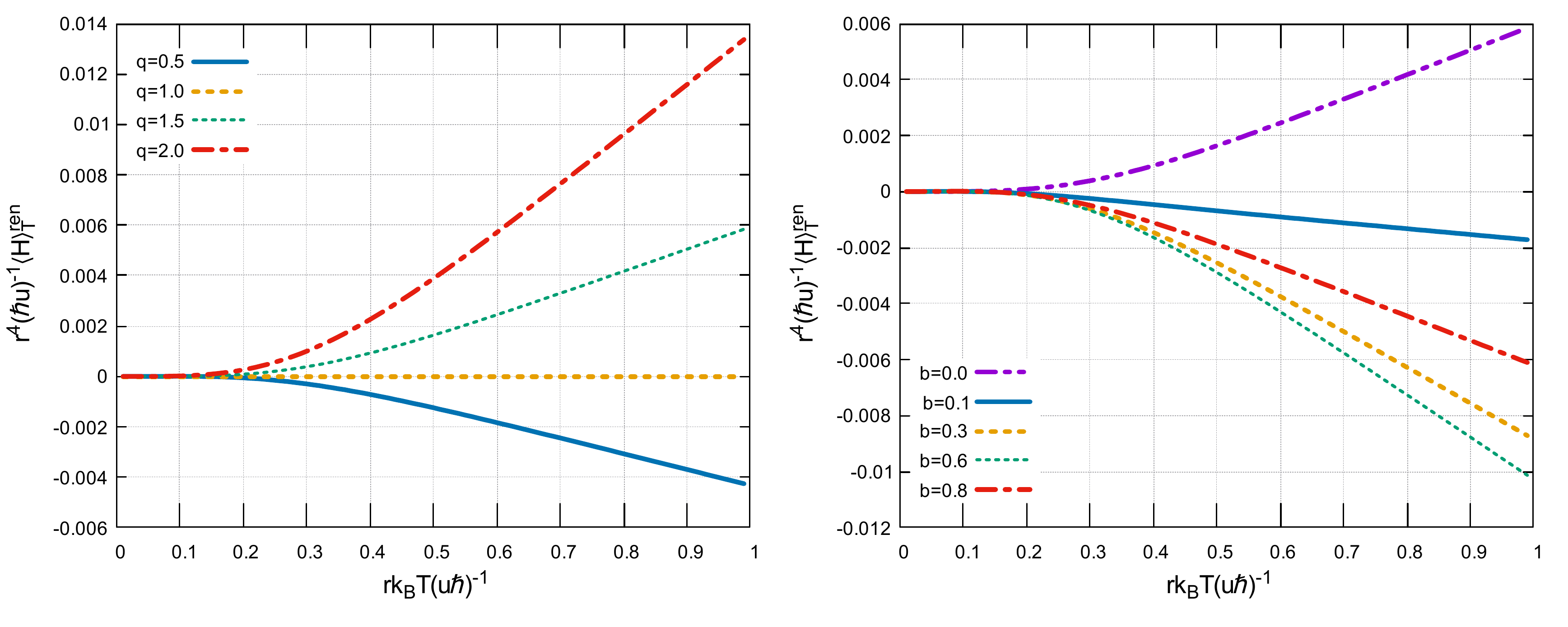}
  \caption{
Renormalized energy density for $b=0$ (left panel), and $q=1.5$ (right panel) as a function of temperature.}
  \label{fig4}
\end{figure}

In order to find a renormalized form for the total energy density in Eq. \eqref{3.11}, we should subtract the black-body radiation contribution in the first term on the r.h.s. Its behavior is shown in Fig.\ref{fig4}. Similar to the free energy density, the total energy density \eqref{3.11} becomes linear in temperature for large $T$, satisfying the classical prediction.

We should now discuss some general considerations about the energies calculated here. All of them are proportional to $r^{-4}$ and consequently, near the origin, $r\to 0$, they diverge. However, for $r\to\infty$ they converge to zero. Note also that, if compared with the Casimir effect, the parameter $b$ from the quasi-periodic boundary condition can control the type of interaction, i.e., it may be repulsive (positive energy), attractive (negative energy), or even null depending on the value of the parameter. This can be seen in the right panel of Fig.\ref{fig4}. This is also true for the conical parameter $q$, which is evident in the left panel of the figure.

\section{Entropy density}
\label{sec4}

Another important thermodynamical quantity to explore is entropy. We find the entropy density taking the derivative of the free energy density given in \eqref{3.6.2}, with respect to the temperature, in the following form \cite{Landau:1980mil}
\begin{eqnarray}
\mathcal{S}&=&-\frac{\partial \mathcal{F}}{\partial T}\nonumber\\
\ &=&\frac{\hslash u}{\pi^2}\sum_{j=1}^{\infty}\left\{\frac{4k_{\rm B}}{(u\hslash j)^4\beta^3}+\frac{4\gamma^3k_{\rm B}}{u\hslash r^3}\left[\sum^{[q/2]}_{n=1}\!^*2\cos(2b\pi n)\eta_j(s_n)-\frac{q}{\pi}\int_0^{\infty}dy M(y,b,q)\eta_j(s_y)\right]\right\},
\label{4.1}
\end{eqnarray}
where
\begin{eqnarray}
\eta_j(s)=\frac{j^2}{(j^2+4\gamma^2s^2)^3}
\, ,
\end{eqnarray}
and we have considered the expression \eqref{3.6.4} for the free energy density to obtain the entropy. Note that the zero-temperature contribution to the free energy density does not contribute to the entropy. Additionally, to obtain a closed-form for the entropy density, we can perform the sum in $j$ by using the following recurrence formula
\begin{equation}
\sum_{j=1}^{\infty}\eta_j(s)=\frac{1}{128\gamma^3s^3}\pi\left(\coth\left(2\pi\gamma s\right)+2 \pi\gamma s\left(1-4\pi\gamma s\coth\left(2\pi\gamma s\right)\right)\text{csch}^2\left(2\pi\gamma s\right)\right).
\label{4.2}
\end{equation}
This leads to
\begin{eqnarray}
\mathcal{S}&=&\frac{2k_B^4\pi^2T^3}{45u^3\hslash^3}+
\frac{k_B}{\pi r^3}\left\{\sum_{n=1}^{[q/2]}\!^*\frac{2\cos(2\pi bn)}{32s_n^3}
\eta(s_n)-\frac{q}{\pi}\int_0^{\infty}dy\frac{M(y,b,q)}{32s_y^3}
\eta(s_y)\right\},
\label{4.3}
\end{eqnarray}
with
\begin{equation}
\eta(s)=\left(\coth\left(2\pi\gamma s\right)+2 \pi\gamma s\left(1-4\pi\gamma s\coth\left(2\pi\gamma s\right)\right)\text{csch}^2\left(2\pi\gamma s\right)\right).
\label{4.3.1}
\end{equation}
Note that by substituting the free energy density \eqref{3.6.2} and the entropy density \eqref{4.1} into the Legendre transform $F=U-TS$, we conclude that the constant $C$ in the total free energy density \eqref{3.6.2} must be null to obtain the total internal energy density \eqref{3.6}.

\begin{figure}[h]
  \includegraphics[scale=0.4]{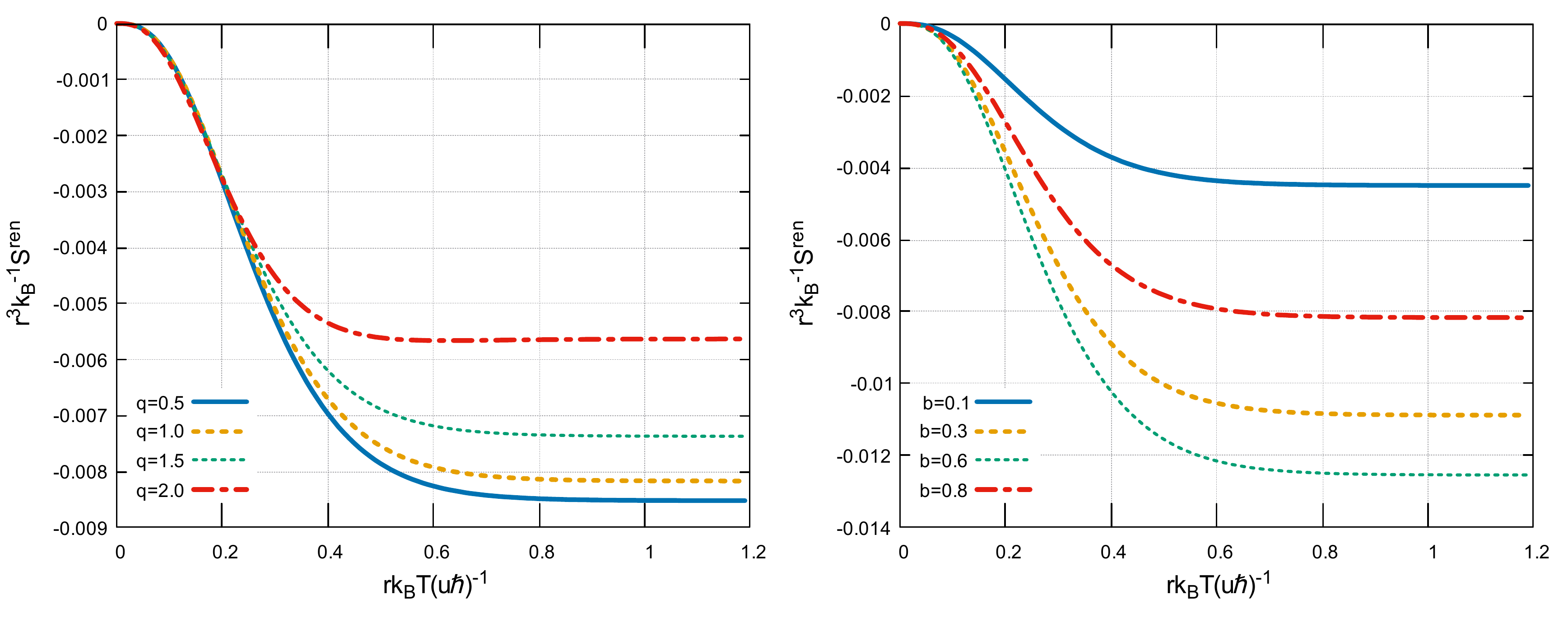}
  \caption{
Entropy density for $b=0.8$ (left panel), and $q=1$ (right panel) as a function of temperature.}
  \label{fig5}
\end{figure}

The renormalized entropy density as a function of temperature for several quasi-periodicity and conical parameters $b$ and $q$ is shown in Fig.\ref{fig5}. This result is due to the field's energy density fluctuations, which characterize an effect analogous to the Casimir effect.
As one can see in the left panel of Fig.\ref{fig5}, the entropy density vanishes when $T\to 0$, regardless of the values of the parameters $b$ and $q$. This result is consistent with the third law of thermodynamics (Nernst heat theorem) \cite{Landau:1980mil,tolman1987relativity}. On the other hand, at high temperatures, the entropy density \eqref{4.3} can be approximated by
\begin{eqnarray}
\mathcal{S}^{\rm ren}\approx\frac{k_B}{32\pi r^3}\left\{\sum_{n=1}^{[q/2]}\!^*\frac{2\cos(2\pi bn)}{s_n^3}-\frac{q}{\pi}\int_0^{\infty}dy\frac{M(y,b,q)}{s_y^3}\right\},
\label{4.4}
\end{eqnarray}
which is temperature independent, in agreement with the classical limit. The behaviors at low and high temperatures noted here can be seen in the plots of Fig.\ref{fig5}. Another point to note is that the renormalized entropy density, \eqref{4.3} without the radiation term, is proportional to $r^{-3}$. Therefore, it converges to zero as $r\to \infty$ and diverges as $r\to 0$ for any case presented here. Although the analog phonon model has been studied in \cite{ford2009fluid,ford2009phononic,deFarias:2021qdg} in several cases, this paper is the first to treat the respective model in a quasi-periodically identified conical spacetime at finite temperature.


\section{Conclusion}\label{sec5}
This paper has investigated the influence of the temperature in a classical liquid described by quantized sound waves whose modes are subject to a quasi-periodic condition, characterized by the parameter $b$, and propagate in the nontrivial topology of a (3+1)-conical spacetime, with conicity parameter $q>0$. The influence of the temperature has also been investigated on relevant physical quantities, such as the mean square density of the liquid and the thermodynamics quantities free energy, internal energy, and entropy densities.
 With the solution of the Klein-Gordon equation given in \eqref{2.5}, the temperature was introduced by the thermal Hadamard function, Eq. \eqref{2.7}, considering the positive and negative frequency solutions.
Under the conditions mentioned above, the solutions were employed to obtain an analytical expression for the thermal Hadamard function \eqref{2.11}, characterizing the explicit dependence on the temperature along with a temperature-independent contribution. Moreover, we have shown that this thermal solution vanishes in the limit $T\to 0$ as it should, leading to the results in the absence of the temperature presented in Ref. \cite{deFarias:2021qdg}.
Thanks to the analytical form of the thermal Hadamard function, the mean square density fluctuation has been obtained for the thermal case \eqref{2.16}. The behavior of the renormalized mean square density fluctuation was shown in Fig.\ref{fig1} where it tends to a constant for  $T\to\infty$, as described by Eq. \eqref{2.16.3}.

Furthermore, the direct relationship between the mean square density fluctuation and internal energy $\mathcal{U}$ has been presented in Eq. \eqref{3.6}, showing that both are the same except for a multiplicative constant. Taking Eq. \eqref{3.6.2}, an analytical expression for the free energy density has been found. The analysis of the renormalized part shows that at high temperatures, it is linear in temperature, matching the result in Fig.\ref{fig3}. This is precisely what is predicted in the classical limit.
It is worth highlighting that the radiation term of both quantities, representing the Minkowski contribution, agrees with the result in \cite{lifshitz2013statistical}. Our results are also consistent with those in \cite{Davies:1987th}, where the results are particular cases of the expressions presented here.
Finally, the total energy density has been obtained by summing the internal energy density \eqref{3.6}, found earlier, with the kinetic energy density one given in \eqref{3.9} providing Eq. \eqref{3.11}.
The behavior of the renormalized part in terms of the temperature may be seen in Fig \ref{fig4}. Thereby, we conclude that the total energy density has a similar behavior if compared with the renormalized free energy density, although with the opposite sign.

Finally, we have found a closed expression for the entropy density, Eq. \eqref{4.3}, which is one of the most important quantities in thermodynamics. We have explored its asymptotic limit $T\to 0$, which goes to zero following the third law of thermodynamics. At high temperatures, the entropy density \eqref{4.3} converges to a constant, consistent with the classical limit. The behavior of the entropy density with temperature can be seen in Fig.\ref{fig5}. We have also verified that when $T\to0$, the thermal part of all physical observables obtained here vanishes, and there remains only the non-thermal contribution obtained in \cite{deFarias:2021qdg}.


{\acknowledgments}
K.E.L.F would like to thank the Brazilian agency CAPES for financial support. AM thanks the financial support from the Brazilian agencies, CAPES, and CNPq under Grant No. 309368/2020-0, and also Universidade Federal de Pernambuco Edital Qualis A. The author H.F.S.M. is supported by the Brazilian agency CNPq under Grants No. 305379/2017-8 and No. 311031/2020-0. 


\end{document}